\documentclass[a4paper,twocolumn,amssymb,final,longbibliography,url,preprintnumbers]{revtex4-1}
\usepackage{hyperref}

\begin{document}
\title{Monopole-antimonopole pair production by magnetic fields}

\author{Arttu Rajantie}
\address{Department of Physics, Imperial College London, London SW7 2AZ, UK}

\email{a.rajantie@imperial.ac.uk}

\begin{abstract}
Quantum electrodynamics predicts that in a strong electric field, electron-positron pairs are produced by the Schwinger process, which can be interpreted as quantum tunnelling through the Coulomb potential barrier. If magnetic monopoles exist,  monopole-antimonopole pairs would be similarly produced in strong magnetic fields by the electromagnetic dual of this process. The production rate can be computed using semiclassical techniques without relying on perturbation theory, and therefore it can be done reliably in spite of the monopoles’ strong coupling to the electromagnetic field. This article explains this phenomenon and discusses the bounds on monopole masses arising from the strongest magnetic fields in the Universe, which are in neutron stars known as magnetars and in heavy ion collision experiments such as lead-lead collisions carried out in November 2018 in the Large Hadron Collider at CERN. It will also discuss open theoretical questions affecting the calculation.\\
\\
{\em Based on a talk given at The Royal Society Scientific Discussion Meeting: Topological Avatars of New Physics, 4-5 March 2019.}
\end{abstract}

\maketitle

\section{Introduction}
Why do electrically charged particles exist but magneti-cally charged particles, i.e., magnetic monopoles, appa-rently do not~\cite{Rajantie:2012xh,Rajantie:2016paj}? If they did, Maxwell's equations would have a perfect duality symmetry between electricity and magnetism.
In 1931, Dirac showed that (static) magnetic monopoles are compatible with quantum mechanics~\cite{Dirac:1931kp}, provided that their magnetic charge $g$ and the electric charges $e$ of all electrically charged particle satisfy the Dirac quantisation condition,
\begin{equation}
\frac{eg}{2\pi}\in{\mathbb Z}.
\label{equ:Diraccondition}
\end{equation}
This implies that both electric and magnetic charges have to be quantised, and if the elementary electric charge is the charge of the positron, then the elementary magnetic charge is the Dirac charge,
\begin{equation}
g_D=\frac{2\pi}{e}\approx 20.7.
\end{equation}

Because $g_D\gg 1$, quantum field theory of magnetic monopoles cannot be studied using standard perturbation theory techniques.
One consequence of this failure of perturbation theory is that it is not possible to compute the production cross section of magnetic monopoles in particle collisions. Collider searches of magnetic monopoles can therefore only report upper bounds on the cross section, rather than actual constraints on the theory parameters such as the monopole mass. In practice, it is customary for the experiments to quote mass bounds obtained by assuming the tree-level Drell-Yan cross section, but these bounds cannot be taken literally and serve mainly as a way of comparing the performance of different experiments.

Furthermore, for solitonic monopoles such as the 't~Hooft-Polyakov monopole~\cite{tHooft:1974kcl,Polyakov:1974ek}, there are semiclassical arguments that their production cross section in elementary particle collisions would be exponentially suppressed~\cite{Witten:1979kh,Drukier:1981fq}. The argument does not apply, at least in the same form, to elementary magnetic monopoles, but it still raises the question of whether their production cross section may also be very different from the Drell-Yan estimates.

In this paper, I will discuss a different monopole production process, namely Schwinger pair production in strong magnetic fields~\cite{Affleck:1981ag,Affleck:1981bma,Gould:2017zwi}. Its rate can be computed using semiclassical instanton techniques, without having to assume a weak coupling. Therefore it does not suffer from the same strong-coupling issue as the perturbative calculations. It is also largely independent of the microscopic details of the monopoles, and within limits, the same results will therefore apply to both elementary and solitonic monopoles. This means that it is possible to obtain actual, largely model-indepenent lower bounds on the magnetic monopole mass.

The strongest known magnetic field in the Universe are in certain neutron stars called magnetars and in heavy ion collision experiments. I discuss the Schwinger process in both cases, and review the monopole mass bounds obtained from them.

\section{Elementary and Solitonic Monopoles}
In principle, magnetic monopoles can appear in quantum field theory as either elementary or composite particles. In the former case, there is a separate quantum field associated with the monopole, whereas in the latter case, they are states made up of other fields. At weak coupling, this state is usually a semiclassical soliton solution, and therefore I will refer to these as solitonic monopoles. 

The best known example of a solitonic monopole is the 't~Hooft-Polyakov monopole solution~\cite{tHooft:1974kcl,Polyakov:1974ek} in the Georgi-Glashow model~\cite{Georgi:1972cj}, an ${\rm SU}(2)$ gauge theory with an adjoint scalar field $\Phi$. When the scalar field has a non-zero vacuum expectation value, it breaks the gauge symmetry to ${\rm U}(1)$, and therefore the low energy effective theory corresponds to electrodynamics.

In the classical theory, the 't~Hooft-Polyakov monopole is a smooth solution of the field equations, of the form
\begin{equation}
\Phi^a(\vec{x})\propto x^a,\quad A_i^a \propto \epsilon_{iaj}x_j,
\end{equation}
and it has the magnetic charge $g=4\pi/e=2g_D$, where $e$ is the ${\rm SU}(2)$ gauge coupling and also corresponds to the elementary electric charge. The monopole mass is given by the energy of the solution,
and is approximately
\begin{equation}
M\approx \frac{4\pi m}{e^2}\approx 137m,
\label{equ:tPmass}
\end{equation}
where $m$ is the mass of the massive gauge bosons. The monopole also has a finite size,
\begin{equation}
R\approx \frac{1}{m} \approx \frac{4\pi}{e^2 M}\approx 137M^{-1}.
\label{equ:tPradius}
\end{equation}

The same monopole exists also in the quantum theory as a non-perturbative state, and at weak coupling it is well approximated by the classical solution. Quantum corrections to it can be calculated perturbatively~\cite{Kiselev:1988gf}, because the coupling constant of the theory is $e\ll 1$, not $g=4\pi/e\gg 1$. One can also go beyond perturbation theory by using lattice Monte Carlo methods~\cite{Rajantie:2005hi,Rajantie:2011nq}.

The 't~Hooft-Polyakov solution can be found in any Grand Unified Theory (GUT)~\cite{Preskill:1979zi}, and therefore the existence of magnetic monopoles is an unavoidable consequence of grand unification.
The mass of these GUT monopoles (\ref{equ:tPmass}) would typically be around
$10^{16}~{\rm GeV}$, which is well above the energies of any foreseeable particle collider experiments. However, there are theories that have lighter solitonic monopole solutions~\cite{Cho:2013vba,Ellis:2016glu}, possibly even within the reach of the Large Hadron Collider.

Although most theoretical work has focused on solitonic monopoles, it is also important to consider the case of elementary monopoles.
In practice, theoretical calculations with them are difficult, not only because of their strong charge, which makes perturbation theory invalid, but also because the existing quantum theory formulations are cumbersome~\cite{Schwinger:1966nj,Zwanziger:1970hk,Milton:2006cp}. However, that is not an argument against their existence.
If the magnetic monopole is an elementary particle, its mass is a free parameter. It is therefore perfectly possible that it is at the TeV scale or even lower, as long as it is compatible with the current experimental bounds.

Because of quantum effects, even elementary monopoles would have to have a non-zero effective size~\cite{goebel1970spatial,goldhaber1982monopoles},
\begin{equation}
R\gtrsim R_{\rm QM}=\frac{g^2}{8\pi M},
\label{equ:quantumradius}
\end{equation}
where $M$ is the monopole mass. This is significantly larger than their Compton radius, and therefore even elementary monopoles would not actually appear fully pointlike. However, because of the calculational difficulties due to the strong magnetic charge, there is no detailed understanding of how this finite size arises. It is, nevertheless, interesting to note that it agrees with the size (\ref{equ:tPradius}) of the 't~Hooft-Polyakov monopole. Indeed, the distinction between elementary and solitonic monopoles is not necessarily clear-cut, and there are examples of theories which have two dual descriptions, in one of which the monopoles are solitonic and in the other one elementary~\cite{Montonen:1977sn}.

\section{Production Cross Section}

There have been several searches for magnetic monopoles in particle colliders~\cite{Patrizii:2015uea}, including LEP, Tevatron and the LHC. The more recent results are from the ATLAS and MoEDAL experiments at the LHC~\cite{Aad:2019pfm,Acharya:2019vtb}. 
Because there has been no positive discovery, these experiments place upper bounds on the monopole production cross section.
In order to constrain actual theory parameters such as the monopole mass, one would need a reliable theoretical prediction for the production cross section. Sadly such a prediction does not exist for monopole production in collisions of elementary particles.

The main obstacle to the calculation of the production cross section of magnetic monopoles is their strong magnetic charge $g=ng_D$, $n\in\mathbb{Z}$, which the Dirac quantisation condition (\ref{equ:Diraccondition}) requires to be much greater than one. This means that the calculation cannot be carried out using perturbation theory. 
There are also strong arguments that the production cross section of solitonic monopoles, such as GUT monopoles, is suppressed by an exponential factor $\exp(-4/\alpha)\sim 10^{-238}$~\cite{Witten:1979kh,Drukier:1981fq}. This is because they are highly ordered coherent lumps of field consisting of $O(1/\alpha)$ quanta. Even if there is enough energy available to produce the monopoles, it is much more likely that the same energy gets distributed to a large number of particles in a less ordered fashion. It is not known if the production cross section of elementary monopoles is suppressed by a similar factor, because we currently do not have the tools to carry out the calculation.

Because of this theoretical uncertainty, experiments tend to quote nominal mass bounds based on the tree-level Drell Yan cross section.
For monopoles with a single Dirac charge, $g=g_D$, this gives~\cite{Aad:2019pfm} $M\gtrsim 1850~{\rm GeV}$ or $M\gtrsim 2370~{\rm GeV}$, depending on whether they have spin $0$ or $1/2$, respectively.
However, these nominal bounds are mainly useful for comparing different experiments and should not be interpreted as actual lower bounds on monopole masses. Much lighter monopoles can exist if their production cross sections is low. In order to obtain an actual mass bound, one therefore has to consider other processes than monopole pair production from elementary particle collisions.

\section{Schwinger Pair Production}
In addition to elementary particle collisions, magnetic monopoles can also be produced in a strong external magnetic field by the Schwinger process. 

It was shown by Sauter~\cite{Sauter:1931zz} and Schwinger~\cite{Schwinger:1951nm} that electrically charged particles are
pair produced in a sufficiently strong electric field. This can be understood as tunneling through the Coulomb potential barrier. The rate of this process can be calculated using semiclassical instanton techniques even at strong coupling~\cite{Affleck:1981ag,Affleck:1981bma}.

If state $|\Omega\rangle$ is unstable, then its decay probability is given by
\begin{eqnarray}
	P&=&1-\left|\langle\Omega|\hat{S}|\Omega\rangle\right|^2
	=1-\exp\left[2\,{\rm Im}\left(i\log \langle\Omega|\hat{S}|\Omega\rangle\right)\right]
	\nonumber\\
		&=&1-\exp\left[2\,{\rm Im}\left(\log \langle\Omega|\hat{S}_E|\Omega\rangle\right)\right],
\end{eqnarray}
where the S-matrix $\hat{S}$ is the time evolution operator from past infinity to future infinity in Minkowski space and $\hat{S}_E$ the corresponding operator in Euclidean space. In the semiclassical approximation, the exponent is given by
\begin{equation}
{\rm Im}\left(\log \langle\Omega|\hat{S}_E|\Omega\rangle\right)\sim e^{-S_{\rm inst}},
\end{equation}
where $S_{\rm inst}$ is the action of the instanton solution, which is a classical solution of the Euclidean equations of motion with one negative mode, i.e. a saddle point solution. One negative mode is needed so that the solution gives an imaginary contribution to the path integral.

For an electrically charged point particle with mass $m$ and electric charge $e$ in a background gauge field $A^{\rm ext}_\mu$, the Euclidean action is
\begin{eqnarray}
S_E[x^\mu]&=&m\int_0^1  d\tau\, \left(\dot{x}_\mu \dot{x}^\mu\right)^{1/2}-ie\oint dx^\mu A^{\rm ext}_\mu
\nonumber\\
&&
-\frac{e^2}{8\pi^2}\int d\tau d\tau'
 \frac{\dot{x}^\mu(\tau) \dot{x}_\mu(\tau')}{|x(\tau)-x(\tau')|^2}
,
\end{eqnarray}
where $\tau\in [0,1)$ is a parameter along the worldline and $A^{\rm ext}_\mu$ is the background gauge field. The last term corresponds to the self-interactions of the particle.

In a constant background electric field $\vec{E}$, rotation symmetry implies that the solution is a circle in the plane defined by the field $\vec{E}$ and the time direction. Denoting the radius of the circle by $r$, its action is
\begin{equation}
S_E(r)=2\pi m r - e|\vec{E}| \pi r^2 - \frac{e^2}{4},
\end{equation}
where the unstable direction corresponds to change of $r$. 
Therefore the saddle point corresponds to the radius
\begin{equation}
r=r_{\rm inst}=\frac{m}{e|\vec{E}|},
\label{equ:instantonradius}
\end{equation}
which maximises the action, and gives
\begin{equation}
S_{\rm inst}=S_E\left(r_{\rm inst}\right)=\frac{\pi m^2}{e|\vec{E}|} - \frac{e^2}{4}.
\end{equation}
The solution has zero modes corresponding to translations in the four Euclidean directions, which contribute a spacetime volume factor, which means that there is a non-zero, finite rate per unit spacetime volume~\cite{Affleck:1981bma}
\begin{equation}
\Gamma= D e^{-S_E}=
\frac{e^2|\vec{E}|^2}{8\pi^3}
\exp\left( - \frac{\pi m^2}{e|\vec{E}|} + \frac{e^2}{4} \right),
\label{equ:GammaE}
\end{equation}
where the prefactor $D$ is given by a functional determinant of the second derivatives of the action.

The result (\ref{equ:GammaE}) means that if the field is sufficiently strong,
\begin{equation}
|\vec{E}|\gtrsim \frac{\pi m_{\rm e}^2}{e}\approx 10^{18}~{\rm V/m},
\end{equation}
where $m_{\rm e}$ is the electron mass,
electron-positron pairs are produced at an unsuppressed rate. This field is a few orders of magnitude stronger than what can be  currently reached with the most powerful lasers, and therefore Schwinger pair production has not yet been confirmed directly in experiments~\cite{Turcu:2016dxm,Gould:2018efv}.

By the electromagnetic duality, if magnetic monopoles exist, they would be pair produced by the same mechanism in sufficiently strong external magnetic field. The rate can be obtained from Eq.~(\ref{equ:GammaE}) by replacing $e\rightarrow g$ and $\vec{E}\rightarrow \vec{B}$~
\cite{Affleck:1981ag,Affleck:1981bma},
\begin{equation}
\Gamma=
\frac{g^2|\vec{B}|^2}{8\pi^3}
\exp\left( - \frac{\pi M^2}{g|\vec{B}|} + \frac{g^2}{4} \right).
\label{equ:GammaB}
\end{equation}
Because $g\gg 1$, the second term in the exponent is important, and therefore the field strength needed to produce monopoles of mass $M$ is
\begin{equation}
|\vec{B}|\gtrsim \frac{4\pi M^2}{g^3}.
\end{equation}
Correspondingly, if monopole production is not observed in field $\vec{B}$, it implies a lower mass bound
\begin{equation}
M\gtrsim \sqrt{\frac{g^3|\vec{B}|}{4\pi}}.
\label{equ:vacuumbound}
\end{equation}

This calculation assumes that the monopoles are pointlike.
The radius of the instanton is given by the electromagnetic dual of Eq.~(\ref{equ:instantonradius}),
\begin{equation}
r_{\rm inst}=\frac{M}{g|\vec{B}|},
\label{equ:instantonradius2}
\end{equation}
so this assumption is justified if the monopole size (\ref{equ:quantumradius}) is less than this,
$R_{\rm QM}\ll r_{\rm inst}$. It is easy to check that this is true if the monopole mass satisfies Eq.~(\ref{equ:vacuumbound}).

As a simple application, one can consider Schwinger pair production of monopoles by the LHC magnets,
which have field strength $|\vec{B}|\approx 8.3~{\rm T}\approx 1.6\times  10^{-15}~{\rm GeV}^2$. Even before any particle collisions were carried out, the fact that this field did not produce magnetic monopoles when the magnets were first switched on, implies the lower mass bound
\begin{equation}
M\gtrsim 1.5\left(\frac{g}{g_D}\right)^{3/2}~{\rm keV}
\label{equ:LHCbound0}
\end{equation}
for the mass of monopoles.

\section{Neutron Stars}
To improve the bound (\ref{equ:LHCbound0}), one needs to find stronger magnetic fields. The strongest known magnetic fields currently existing in the Universe and in neutron stars known as magnetars, 23 of which have been found~\cite{Olausen:2013bpa}. 
The one with the strongest field is SGR 1806-20, with $|\vec{B}|\approx 2\times 10^{11}~{\rm T}\approx 4\times 10^{-5}~{\rm GeV}^2$.
Its temperature is low in comparison, $T\approx 0.55~{\rm keV}$, and therefore the Schwinger pair production rate is given by
the zero-temperature expression~(\ref{equ:GammaB}).

At the surface of SGR 1806-20, the ratio of the gravitational and magnetic forces acting on a monopole
is
\begin{equation}
\frac{F_G}{F_B}=\frac{G_N M_{\rm NS} M}{g|\vec{B}|R_{\rm NS}^2}
\approx
\left(
\frac{g}{g_D}\right)^{-1}
\frac{M}{1.8\times 10^{17}~{\rm GeV}},
\end{equation}
where $R_{\rm NS}\sim 10~{\rm km} \approx  5\times 10^{19}~{\rm GeV}^{-1}$ is the radius of the magnetar and $M_{\rm NS}\sim 1.5M_{\odot}\approx 1.6\times 10^{57}~{\rm GeV} $ is its mass. 
If a pair of magnetic monopoles with mass $M\ll 1.8\times 10^{17}~{\rm GeV}$ are produced near the surface of a magnetar, the magnetic field would therefore pull one of them to the surface of the star and expel the  other one into space. This would reduce the strength of the magnetic field, in contradiction with observations~\cite{Gould:2017zwi}. Using Eq.~(\ref{equ:vacuumbound}), one therefore obtains a bound
\begin{equation}
M\gtrsim 0.17\left(\frac{g}{g_D}\right)^{3/2}~{\rm GeV}.
\label{equ:NSbound}
\end{equation}
A more detailed calculation, which takes into account the long time scale over which the field has to survive and grow, gives a somewhat stronger bound $M\gtrsim 0.31~{\rm GeV}$ for $g=g_D$~\cite{Gould:2017zwi}.

The bound (\ref{equ:NSbound}) was obtained by considering the magnetic field outside the magnetar, but the magnetic fields in the interior are even stronger.
There are also many other open questions about magnetars,
and with a better understanding of them, one may well be able to improve the bound further.

\section{SPS Heavy Ion Collisions}
Even stronger magnetic fields than those around magnetars are present in relativistic heavy ion collisions. The heavy ions are nuclei of heavy elements such as Au or Pb, and therefore they have a high electric charge $Q=Ze\sim 100e$.  In these experiments, these nuclei are collided at relativistic speeds. When the collision is not head-on, one therefore has two very high electric currents moving in opposite directions past each other at the time of the collision. This induces a very strong magnetic field for a short period of time.

The most recent published monopole search in heavy ion collisions was carried out at CERN Super Proton Synchrotron (SPS) in 1997~\cite{He:1997pj}.
It was a fixed-target Pb collision with beam energy $160A~{\rm GeV}$, corresponding to a centre-of-mass energy
$\sqrt{s_{NN}}\approx 17~{\rm GeV}$ per nucleon. This produces a magnetic field 
$|\vec{B}|\approx 0.01~{\rm GeV}^2$~\cite{Skokov:2009qp} and a fireball with a high temperature~$T\approx 0.185~{\rm GeV}$~\cite{Schlagheck:1999aq}. 

The Schwinger process at non-zero temperature was studied in Ref.~\cite{Gould:2017fve,Gould:2018ovk}. 
To calculate the rate in the semiclassical approximation, one needs to find the instanton in Euclidean space with a compact imaginary time direction of length $\beta=1/T$. This breaks the Euclidean rotation symmetry and therefore the instanton is no longer a circle. Because of the strong coupling, the solutions have to be found numerically, and this was done in Ref.~\cite{Gould:2017fve}. The result for the SPS case is, however, simple. Because the temperature is sufficiently high, the relevant instanton is a time-dependent sphaleron solution, which corresponds to a static monopole-antimonopole pair. 

The energy of a static monopole-antimonopole pair separated by the distance $\vec{r}$ in a constant magnetic field $\vec{B}$ is
\begin{equation}
E(\vec{r})=2M-\frac{g^2}{4\pi |\vec{r}|}-g\vec{B}\cdot\vec{r}.
\end{equation}
The sphaleron configuration corresponds to the distance
\begin{equation}
|\vec{r}|=r_{\rm sph}=\sqrt{\frac{g}{4\pi |\vec{B}|}},
\end{equation}
which maximises the energy, which gives the sphaleron energy
\begin{equation}
E_{\rm sph}
=E(r_{\rm sph})=2M-2\left(\frac{g^3|\vec{B}|}{4\pi}\right)^{1/2}.
\end{equation}
Including the prefactor,
the pair production rate is~\cite{Gould:2018ovk}
\begin{equation}
\Gamma\approx
\left(\frac{M^5 T^9}{64\pi^7 g B^3}\right)^{1/2}\exp\left[-\frac{2M}{T}\left(1-\sqrt{\frac{g^3|\vec{B}|}{4\pi M^2}}\right)\right],
\end{equation}
and the predicted monopole pair production cross section is
\begin{equation}
\sigma_{M\overline{M}}\approx \sigma_{\rm tot}{\cal V}\Gamma,
\end{equation}
where $\sigma_{\rm tot}\approx 6.3~{\rm b}$ is the total inelastic cross section, and ${\cal V}$ is the spacetime volume of the collision.
The failure of SPS heavy ion collisions to produce magnetic monopoles implies an upper bound 
$\sigma_{M\overline{M}}\lesssim 1.9~{\rm nb}$ on the monopole pair production cross section~\cite{He:1997pj}.
This translates to a bound on the monopole mass~\cite{Gould:2017zwi},
\begin{equation}
M\gtrsim
\left(2.0+2.6\left(\frac{g}{g_D}\right)^{3/2}\right)~{\rm GeV}.
\label{equ:SPSbound}
\end{equation}

\section{LHC Heavy Ion Collisions}
The magnetic field produced by a heavy ion collision increases with the collision energy, and therefore the Relativistic Heavy Ion Collider (RHIC) and the Large Hadron Collider (LHC) should be able to provide much stronger bounds on monopole pair production and, therefore, on the monopole mass. There is no published data on monopole searches at RHIC, so I will focus on the LHC, which carried out a month-long heavy ion run in November 2018, with collision energy per nucleon
$\sqrt{s_{NN}}=5.02~{\rm TeV}$.

The evolution of the electromagnetic fields during the collision was studied in Ref.~\cite{Deng:2012pc}. The field strength grows linearly with collision energy, and is highest for collisions with impact parameter $b\approx 13~{\rm fm}$, which corresponds to the diameter of the nucleus. Therefore we are mainly interested in peripheral collisions, and can ignore thermal effects.

For the LHC energy, the peak magnetic field strength is
$B\approx 7.3~{\rm GeV}^2$, reached at the time of collision. The field is highly time-dependent, and can be approximately described by the analytic fit~\cite{Gould:2019myj}
\begin{eqnarray}
\vec{B}&=&\frac{B\hat{y}}{2}\left[
\left(1+\omega^2\left(t-\frac{z}{v}\right)^2\right)^{-3/2}
\right.
\nonumber\\&&
\left.
+
\left(1+\omega^2\left(t+\frac{z}{v}\right)^2\right)^{-3/2}
\right],
\label{equ:timedepB}
\end{eqnarray}
where
$\hat{y}$ is the direction of the impact parameter, and
$\omega\approx 73~{\rm GeV}$ parameterises the rate of the field evolution in time.
The time-dependence is important if the
the ratio of this to the inverse radius (\ref{equ:instantonradius2})) of the constant-field instanton,
\begin{equation}
\xi=\omega r_{\rm inst}=\frac{\omega M}{gB}
\approx \left(\frac{g}{g_D}\right)^{-1}\frac{M}{2~{\rm GeV}},
\end{equation}
is large. Therefore
we see that time dependence cannot be ignored for monopoles heavier than a few ${\rm GeV}$.

In Ref.~\cite{Gould:2019myj}, the pair production rate was computed
for the time-dependent field (\ref{equ:timedepB}), by analytically continuing it to Euclidean time $\tau$, 
\begin{eqnarray}
\vec{B}\longrightarrow \vec{B}_E&=&\frac{B\hat{y}}{2i}\left[
\left(1+\omega^2\left(i\tau-\frac{z}{v}\right)^2\right)^{-3/2}
\right.\nonumber\\&&\left.
+
\left(1+\omega^2\left(i\tau+\frac{z}{v}\right)^2\right)^{-3/2}
\right],
\label{equ:timedepBE}
\end{eqnarray}
and then finding the instanton solution in this Euclidean background field.

In the zeroth-order approximation, in which the monopole worldline self-interactions are ignored completely,
the instanton solution can be found analytically~\cite{Dunne:2005sx,Dunne:2006st,Gould:2019myj} to be an ellipse with semi-major axis $a_\tau=M/gB\sqrt{1+\xi^2}$ in the Euclidean time direction and semi-minor axis
$a_y=M/gB(1+\xi^2)$ in the $y$ direction.
Its action is~\cite{Gould:2019myj}
\begin{equation}
S_E^{(0)}=\frac{4M^2}{gB\xi^2}
\left[
{\bf E}(-\xi^2)
-{\bf K}(-\xi^2)
\right]\sim \frac{4M^2}{gB\xi}
,
\label{equ:SEzero}
\end{equation}
where ${\bf E}$ and ${\bf K}$ are elliptic integrals, and the last expression is valid at $\xi\gg 1$, which corresponds to $M\gg 1~{\rm GeV}$. At the zeroth order, the prefactor in the rate $\Gamma$ can also be computed analytically~\cite{Dunne:2005sx,Dunne:2006st}. 
Importantly, the result (\ref{equ:SEzero}) shows that the rate decreases when $\xi$ increases.

One can also go beyond the zeroth-order approximation analytically, and compute the next-to-leading order self-interaction correction to 
the instanton action~\cite{Gould:2019myj},
\begin{equation}
S_E^{(1)}=-\frac{g^2}{8}\left(
\sqrt{1+\xi^2}
+\frac{1}{\sqrt{1+\xi^2}}
\right)
\sim -\frac{g^2}{8}\xi
.
\label{equ:SEone}
\end{equation}
Comparing Eq.~(\ref{equ:SEzero}) and (\ref{equ:SEone}), we see that
the total next-to-leading order action $S_E^{(0)}+S_E^{(1)}$ is negative if
$\xi^2>32M^2/g^3B$, which is always satisfied for LHC collisions, irrespective of the monopole mass $M$.
At the face value this would imply that the pair production is unsuppressed, but in practice it is more likely to be a signal that the approximations have broken down.

It is possible to find the instanton to all orders in self-interactions by solving the full non-linear equations numerically~\cite{Gould:2019myj}. Because the equations are non-local, this is a computationally heavy task, and therefore it has currently only been done for relatively small $\xi$, where the NLO approximation should be valid, and indeed, the results agree with it.

Again, this calculation is based on the assumption of a pointlike monopole. 
For it to be valid, the monopole size (\ref{equ:quantumradius}) needs to be smaller than the semi-minor axis $a_y$ of the instanton, which gives the condition $\xi^2\lesssim 8\pi M^2/g^3 B$. This coincides approximately with the value where the NLO action in the pointlike approximation becomes negative, and is not satisfied for any monopole masses in LHC heavy ion collisions. For an accurate and reliable estimate of the production rate, it will therefore be necessary to include the non-zero monopole size.
For 't~Hooft-Polyakov monopoles this can be done, at least in principle, by finding the instanton solution in the full field theory.

As a rough estimate of the production cross section, one can use the locally constant field approximation, in which the constant-field rate (\ref{equ:GammaB}) is integrated over the spacetime volume were the fields are strongest. 
Within this approximation,
if the LHC searches do not find monopoles,  Eq.~(\ref{equ:vacuumbound}) would give a very rough bound
\begin{equation}
M\gtrsim 70~{\rm GeV}.
\label{equ:LHCbound}
\end{equation}
Because the effect of the time-dependence appears to enhance the production, this can be expected to be a conservative estimate. A complete calculation may therefore make the bound significantly stronger.

\section{Conclusion}
The electromagnetic dual of
Schwinger pair production provides a new way of searching for magnetic monopoles.
If monopoles exist, they would be produced in a sufficiently strong magnetic field. Conversely, if this does not happen, monopoles either do not exist or their mass is too high. By considering physical instances of strong magnetic fields, we can therefore derive lower bounds on magnetic monopole masses. This requires a theoretical calculation of the pair production rate, which can be carried out using the semiclassical instanton method, which does not require perturbation theory or the assumption of a weak coupling. Therefore it can be applied to magnetic monopoles, whose coupling to the electromagnetic field is necessarily strong. 

In a constant field at zero temperature, the rate can be computed analytically and the result is independent of the microscopic details of the monopoles and whether they are elementary and solitonic. The resulting mass bounds are therefore universal. In a time-dependent field, the full result requires a numerical calculation, and the finite monopole size needs to be taken into account. Therefore the precise result will also depend on the microscopic nature of the monopoles. For solitonic 't~Hooft-Polyakov monopoles, the numerical calculation is straightforward in principle, although computationally very demanding.

Using this approach, the magnetic fields of magnetars imply a bound~(\ref{equ:NSbound}) of slightly below one ${\rm GeV}$, and heavy ion collisions at SPS an order of magnitude stronger~(\ref{equ:SPSbound}). The LHC should be able to improve significantly on these. The results of the one-month heavy-ion run in November 2018 have not yet been published, but the estimate (\ref{equ:LHCbound}) suggests that if monopoles with mass $M\lesssim 70~{\rm GeV}$ exist, they would have been produced then. If data shows no monopoles, it will therefore imply a lower mass bound of the same order. However, obtaining the precise mass bound will need further theoretical work in order to take the time dependence of the collision and the finite monopole size into account.

At the face value, this appears to be $20-30$ times lower than the current bounds from ATLAS and MoEDAL. However, it is important to remember that all existing mass bounds from proton-proton collisions are based on perturbation theory, which is not actually valid for magnetic monopoles because of their strong magnetic charge. Therefore one cannot currently rule out the existence of monopoles with masses of even a few ${\rm GeV}$, and the only way to do that is to carry out these calculations and experiments.

\acknowledgments
The author would like to thank O.~Gould, D.L.-J.~Ho, S.~Mangles, S.~Rose, D.J.~Weir and C.~Xie and the whole MoEDAL collaboration for useful discussions and collaboration on this topic. This work was supported by the UK Science and Technology Facilities Council grant ST/P000762/1.

%


\end{document}